\documentclass[twocolumn,prl,showpacs,superscriptaddress]{revtex4}

\usepackage{graphicx}
\usepackage{times}
\usepackage{rotating}
\usepackage{amsmath}
\usepackage{amsfonts}
\usepackage{amssymb}
\usepackage{enumerate}
\usepackage{longtable}
\usepackage{dcolumn}
\usepackage{bm}
\usepackage{color}
\usepackage[pdftex,colorlinks,bookmarks=false,citecolor=blue,linkcolor=red,urlcolor=blue]{hyperref}

\begin{document}
\newcommand{\cred}{\color{red}}
\title{Solids and supersolids of three-body interacting
polar molecules in an optical lattice}

\author{Kai P. Schmidt}
\email{schmidt@fkt.physik.uni-dortmund.de}
\affiliation{Lehrstuhl f\"ur theoretische Physik I, Otto-Hahn-Str. 4, TU Dortmund, D-44221 Dortmund, Germany}

\author{Julien Dorier}
\affiliation{Institute of Theoretical Physics, \'{E}cole Polytechnique F\'{e}d\'{e}rale de Lausanne, CH 1015 Lausanne, Switzerland}

\author{Andreas M. L\"auchli}
\email{laeuchli@comp-phys.org}
\affiliation{Institut Romand de Recherche Num\'erique en Physique des Mat\'eriaux (IRRMA), CH-1015 Lausanne, Switzerland}
\altaffiliation{Present address: Max Planck Institut f\"ur Physik Komplexer Systeme, N\"othnitzer Str. 38, D-01187 Dresden, Germany}

\date{\rm\today}

\begin{abstract} 
We study the physics of cold polar molecules loaded into an optical lattice in the regime of strong three-body interactions, as put
forward recently by B\"uchler {\em et al.} [Nat. Phys. {\bf 3}, 726 (2007)]. To this end quantum Monte Carlo simulations, exact diagonalization and a semiclassical approach are used to explore hardcore bosons on the 2d square lattice which interact solely by long ranged three-body terms. The resulting phase diagram shows a sequence of solid and supersolid phases. Our findings are directly relevant for future experimental implementations and open a new route towards the discovery of a lattice supersolid phase in experiment.
\end{abstract}


\maketitle
    
\paragraph{Introduction} Strongly correlated systems studied in condensed matter physics or in atomic physics are usually dominated by two-body interactions. The paradigm models are the fermionic and the bosonic Hubbard model which include a local two-body density-density interaction or the Heisenberg model consisting of two-body spin exchanges. These standard models are able to describe an enormous number of physical phenomena since the simultaneous interaction between more than two particles is small in most cases because it arises only in higher order of perturbation theory. Nevertheless multi-body interactions are present and can have profound effects on the physics of a system, e.g. ring-exchange processes being responsible for the rich nuclear magnetism of Helium~3~\cite{roger83}, the accurate description of undoped high-T$_{\rm c}$ compounds and cuprate ladders requires four-spin
interactions \cite{colde01,eccle98,nunne02,notbo07}, while three-body (3B) exchanges appear naturally in the context of two atomic species in a frustrated optical lattice topology~\cite{pacho04a}.\\
On the theoretical side the study of microscopic models with multi-particle interactions is a very active and fruitful line of research. 
Exotic quantum ground states and deconfined criticality~\cite{senthil04} can possibly be triggered by such interactions~\cite{RingNematic,sandvik07,melko08}. Furthermore one can expect fractionalization of elementary excitations, e.g. spin liquid states in quantum magnets~\cite{DimerModels}, topological ordered states as discussed in the context of quantum computation~\cite{kitae06} or fractional 
quantum Hall states~\cite{moore91,fradk98,coope04}. 
The major obstacle on the way towards an experimental confirmation of these fascinating predictions is usually the requirement of dominating multi-body interactions, which is 
hard to achieve in a condensed matter setting.
The field of ultracold gases loaded into optical lattices opens now a new perspective to overcome these difficulties. It has recently been shown that ultracold gases of polar molecules confined to optical lattices can be tuned to a regime where the interactions are solely of 3B type~\cite{buchl07}. 
In contrast to conventional Hubbard or Heisenberg models having mostly short-range interactions, the 3B density interactions put forward in~\cite{buchl07} 
decay only slowly in space, owing to the underlying dipolar interactions. 
It is an important task to study systems based on these novel multi-body interactions and to uncover the nature of unconventional phases harbored in their phase
diagram.\\
\begin{figure}
    \begin{center}
        \includegraphics*[width=0.6\columnwidth]{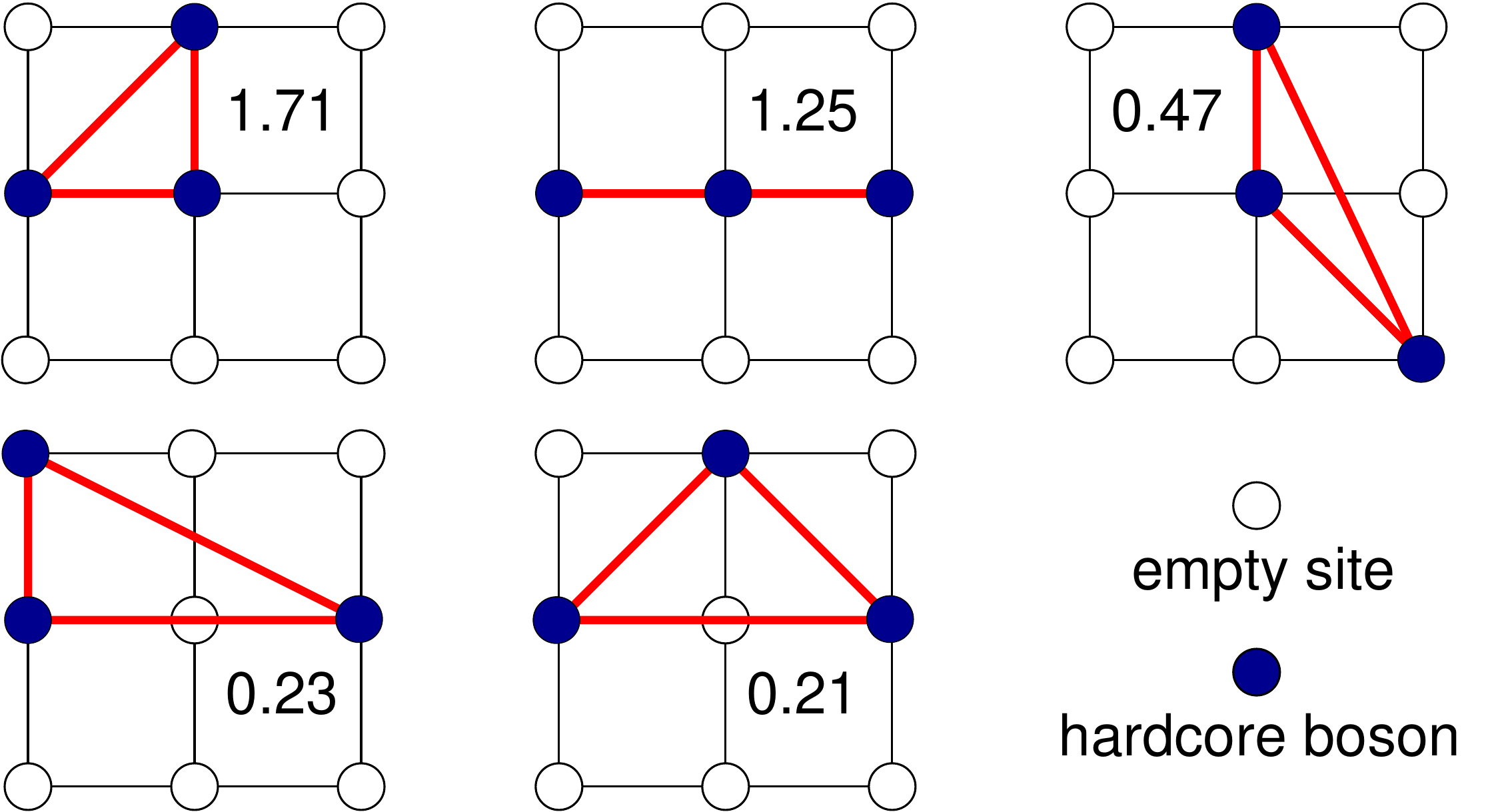}
    \end{center}
    \caption{(Color online) Illustration of the leading 3B interactions $W_{ijk}$ with amplitude larger than $0.2$, defining the minimal model
    studied here.
    }
    \label{fig:sketch}
\end{figure}
\begin{figure*}
    \begin{center}
        \includegraphics*[width=0.7\linewidth]{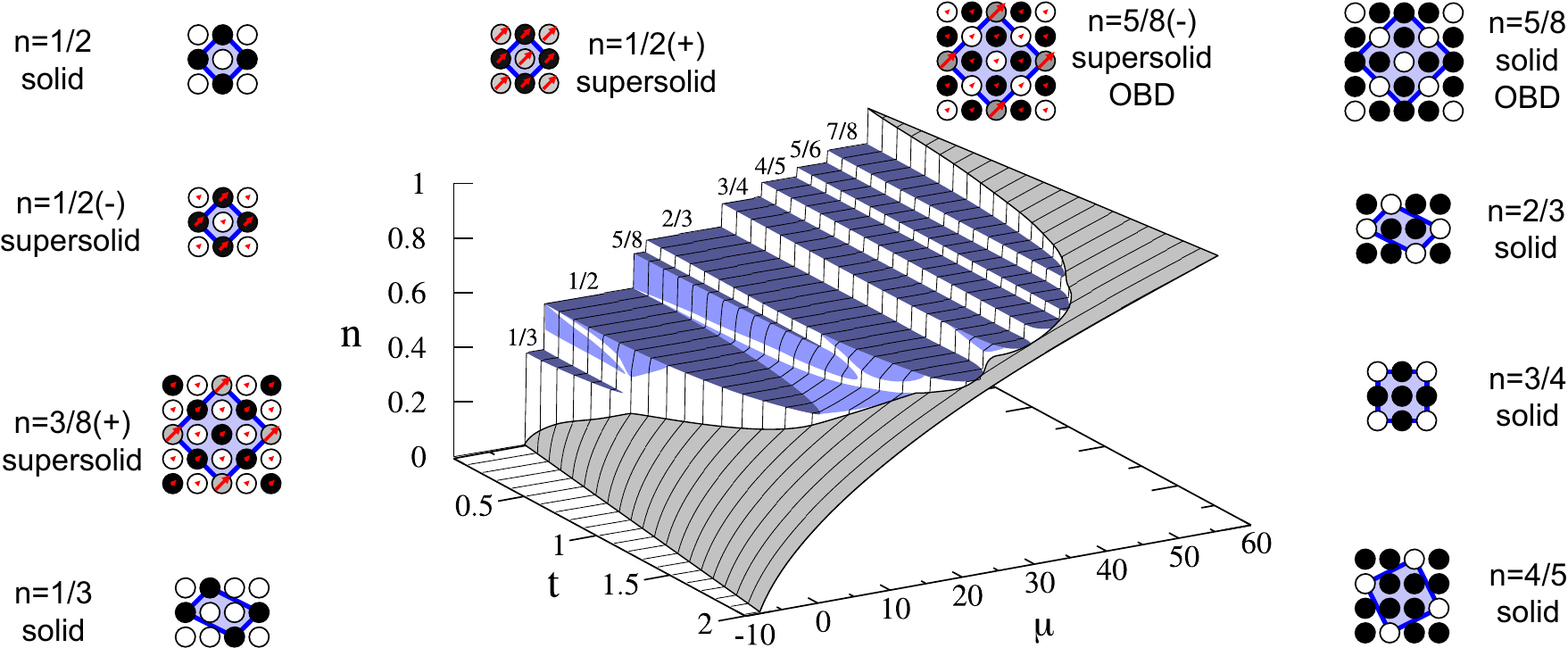}
    \end{center}
    \caption{(Color online) 
    Central plot: Phase diagram of the minimal model obtained within the semiclassical approach as a function of the chemical potential $\mu$ 
    and the nearest-neighbor hopping amplitude $t$.  Grey (white) regions are superfluid (phase separated), and light (dark) blue denotes 
    supersolids (solids). Surrounding plots: schematic representation of the nature of some of the solid and supersolid phases. The greyscale of the
    circles represents the filling (white = empty, black = full), while the length and the direction of the red arrows denotes the amplitude and the phase
    of the superfluid component. The blue lines highlight the unit cell of the different structures.
    }
\label{fig:pd_truncated}
\end{figure*}
In the present Letter, we achieve a step in this direction by exploring the potentially most relevant case of hardcore bosons on a square lattice interacting only 
through 3B forces. We perform a comprehensive numerical study, based on a semiclassical approximation (SCA), exact diagonalization (ED)  
and quantum Monte Carlo simulations (QMC) to derive the resulting zero temperature phase diagram. We reveal a rich sequence of solids, supersolids and phase separation
as the density $n$ is tuned from 0 to 1. Interestingly we find a stable, extended checkerboard supersolid (CSS) phase around density $n=1/2$. Such lattice supersolids are currently a topic of great interest both in the fields of cold atoms and quantum magnetism (see e.g. Refs.~\onlinecite{goral02,schmi08b}).

\paragraph{Model} We consider hard-core bosons hopping on the two-dimensional square lattice including 3B density interactions as put forward in~\cite{buchl07}:
\begin{equation}
 H=-t\sum_{\langle i,j\rangle} (b^\dagger_{i}b^{\phantom{\dagger}}_j +h.c.)-\mu\sum_i n_i+\frac{1}{6}\sum_{i\neq j\neq k}W_{ijk} n_i n_j n_k
 \label{eq:hamilton}
\end{equation}
where $n_i=b^\dagger_i b^{\phantom{\dagger}}_i$ is the boson density at site $i$, $\mu$ is the chemical potential, $t$ is the nearest-neighbor hopping amplitude, and $W_{ijk}$ labels the 3B interactions. \\
The 3B interactions derive from the dipolar forces between the polar molecules under the additional influence of microwave fields~\cite{buchl07},
and retain some of its character, especially the long range nature. The general expression for the amplitudes $W_{ijk}$ is given by:
\begin{equation}
 W_{ijk}=\bar{W}_0 \left[ \frac{1}{|R_i - R_j|^3|R_i - R_k |^3}+\text{permutations}\right]\quad .
\label{eq:full_W}
\end{equation}
The coefficient $\bar{W}_0$ depends on the microscopic setup and is discussed in Ref.~\onlinecite{buchl07}. In the following the energy scale is set by
the 3B interactions and we thus put $\bar{W_0}=1$.  Note that the spatial dependence of the interactions is such that the repulsion between 3 particles 
with a mutual distance of order $R$  amounts only to $1/R^6$. If however two particles are close, while the third is at distance $R$, then the 
interactions only decay as $1/R^3$, resembling the decay of the underlying dipolar interactions. 
Based on these considerations we expect the 3B interactions to have a stronger effect at high densities than at very low density. Furthermore by 
applying a particle-hole transformation in the regime of densities close to $n=1$ one can effectively map the problem to a low-density, two-body 
dipolar gas of holes. So the most challenging regime remains for densities $n\sim 1/2$, where the full structure of the 3B interactions is important.\\
The plan of the paper is to study first a minimal model where the range of the 3B interactions is limited to the 5 types of terms illustrated in Fig.~\ref{fig:sketch}
(i.e. $W_{ijk}>0.2$).
We map out the phase diagram using a semiclassical approach, and confirm the main findings for selected parameters by numerical ED and 
QMC simulations. Then we corroborate the utility of the minimal model by including all terms with amplitudes $W_{ijk}>10^{-3}$ in the 
semiclassical approach.

\paragraph{Semiclassical approximation (SCA)}
The SCA maps Eq.~\ref{eq:hamilton} to a spin 1/2 model using the exact Matsubara-Matsuda~\cite{matsu56} representation of hardcore bosons $S^+=b$, $S^-=b^\dagger$, and $S^z=1/2-n$. The resulting spin Hamiltonian is studied in the classical limit by replacing the quantum spins by classical vectors of length 1/2 on a sphere. The classical ground state is obtained in the thermodynamic limit by numerically determining the global energy minimum among {\em all} non-equivalent unit cells with up to 32 sites. The spin structures minimizing the energy are
mapped back to the boson problem and correspond typically to superfluid, solid or supersolid phases of varying spatial complexity. A supersolid is a phase breaking simultaneously the U(1) gauge symmetry (superfluid) and the underlying translational symmetry of the lattice (solid). The method is computationally much
less expensive than the ED or QMC simulations, and can therefore be used to efficiently map out the phase diagram. \\
The resulting SCA phase diagram for the minimal model as a function of $t$ and $\mu$ is shown in Fig.~\ref{fig:pd_truncated}. 
For large $t$, the system corresponds to basically non-interacting hardcore bosons and is thus expected to be superfluid
for all densities. 
In the opposite limit $t=0$ only commensurate solid phases are found. The density $n(\mu)$ displays a simple series of plateaux which are separated by first order transitions, i.e.
jumps in the density. 
We find plateaux at $n=1/3$, $1/2$, $5/8$, $2/3$, $3/4$, $4/5$, $5/6$, and $7/8$ for the minimal model. 
The much richer structure above $n=1/2$ compared to low densities is a consequence of the particle-hole asymmetry discussed above.
The specific structure of some of the plateaux are illustrated in Fig.~\ref{fig:pd_truncated}. 
Due to the finite range of the truncated interactions the plateaux at $5/8, 5/6$ and $7/8$ exhibit a residual degeneracy in the limit $t=0$, 
which is expected to be lifted either by the longer range couplings (see below) or an order by disorder mechanism driven by the quantum 
fluctuations at finite $t$, which is however beyond the present SCA approach.
Within the SCA the quantum melting of the various solids takes place for values $t\lesssim 2$. 
The physics below $n=1/2$ consists of a superfluid developing from low densities as $t$ increases,
a solid at $n=1/3$, which is destroyed by a rather small amount of hopping, as well as a puzzling
supersolid without corresponding $n=3/8$ plateau~\cite{schmi08b}. Finally the checkerboard solid (CS)
at $n=1/2$ extends to densities below 1/2 by forming a stable CSS with a 
non-trivial dependence on $t$, leading to a maximal region of stability around $t\approx 0.6$. For values
$t\gtrsim0.8$ the transition from the superfluid below to the CS is direct and first order.
The physics above the CS is even richer. Above $n=1/2$, we find a CSS in a large range of $t$ values. 
The plateau at $5/8$ also has a corresponding supersolid 
for a density range below $5/8$, but this supersolid is very compressible. Furthermore we find supersolids
for densities just below $n=2/3$ and $3/4$ in a small window of $t$.\\
The SCA phase diagram of the minimal model is very rich, including several supersolids of different 
spatial struture. We now proceed to a numerically exact treatment of the minimal model to confirm the main findings, 
i.e.~the basic solids and the $n=1/2$ supersolids.
\paragraph{Numerical simulations}
In the following we use QMC and ED in order to corroborate the predictions made by the SCA. We focus first on the case $t=0.5$ displaying most of
the features in the SCA calculation and briefly comment on results obtained for other $t$ values. 
The ED calculations were performed on square clusters up to 36 sites. 
The QMC simulations are based on a modified~\cite{schmi06} stochastic series expansion (SSE)~\cite{sandvik,alet05} code  and the 
ALPS libraries~\cite{alps}.
We restrict ourselves to the density range $0 \leq n \leq 5/8$ for $t=0.5$. On the one hand because the SSE algorithm 
based on the directed loop update is not particularly efficient in exploring the solid phases with large unit cells found at higher densities,
and on the other hand it is also the most interesting density range since it displays a sizable CSS phase in the SCA approach. 
The lowest temperature was typically $T=0.05$, which is representative of the ground state for the shown quantities. Systems sizes went 
up to $N=12\times 12=144$ sites.
\begin{figure}
        \includegraphics*[width=0.76\columnwidth]{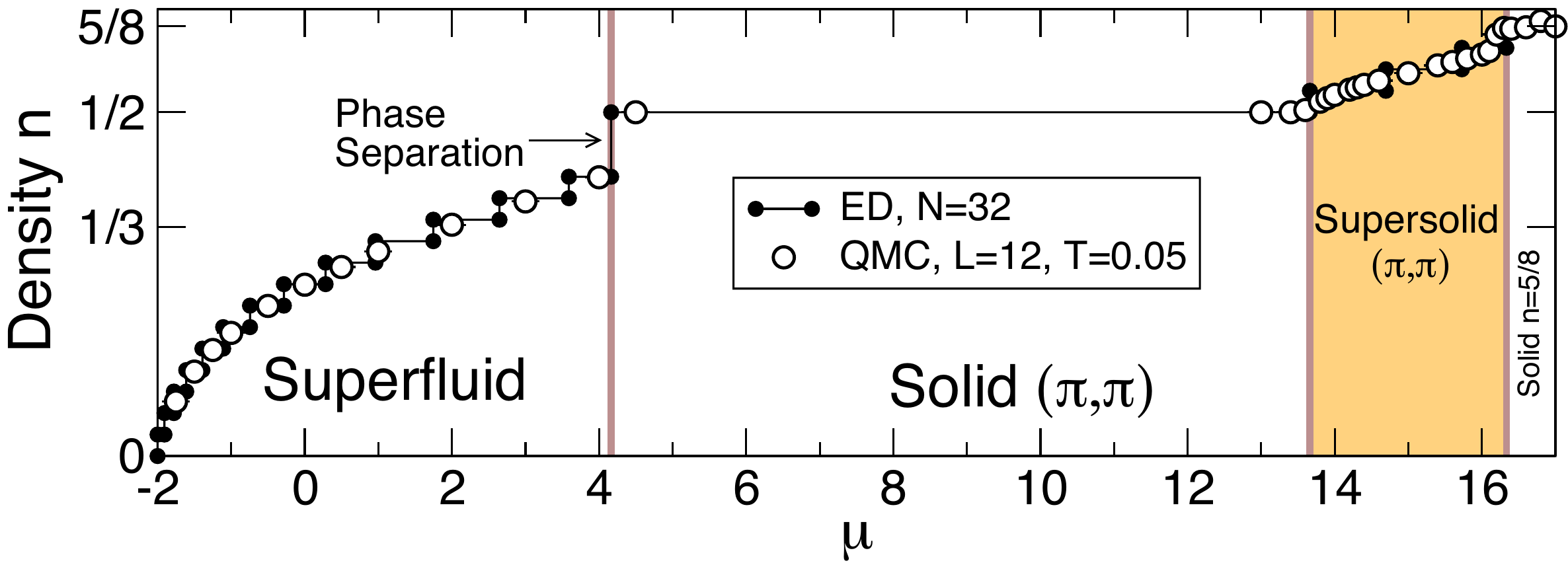}
        \includegraphics*[width=0.78\columnwidth]{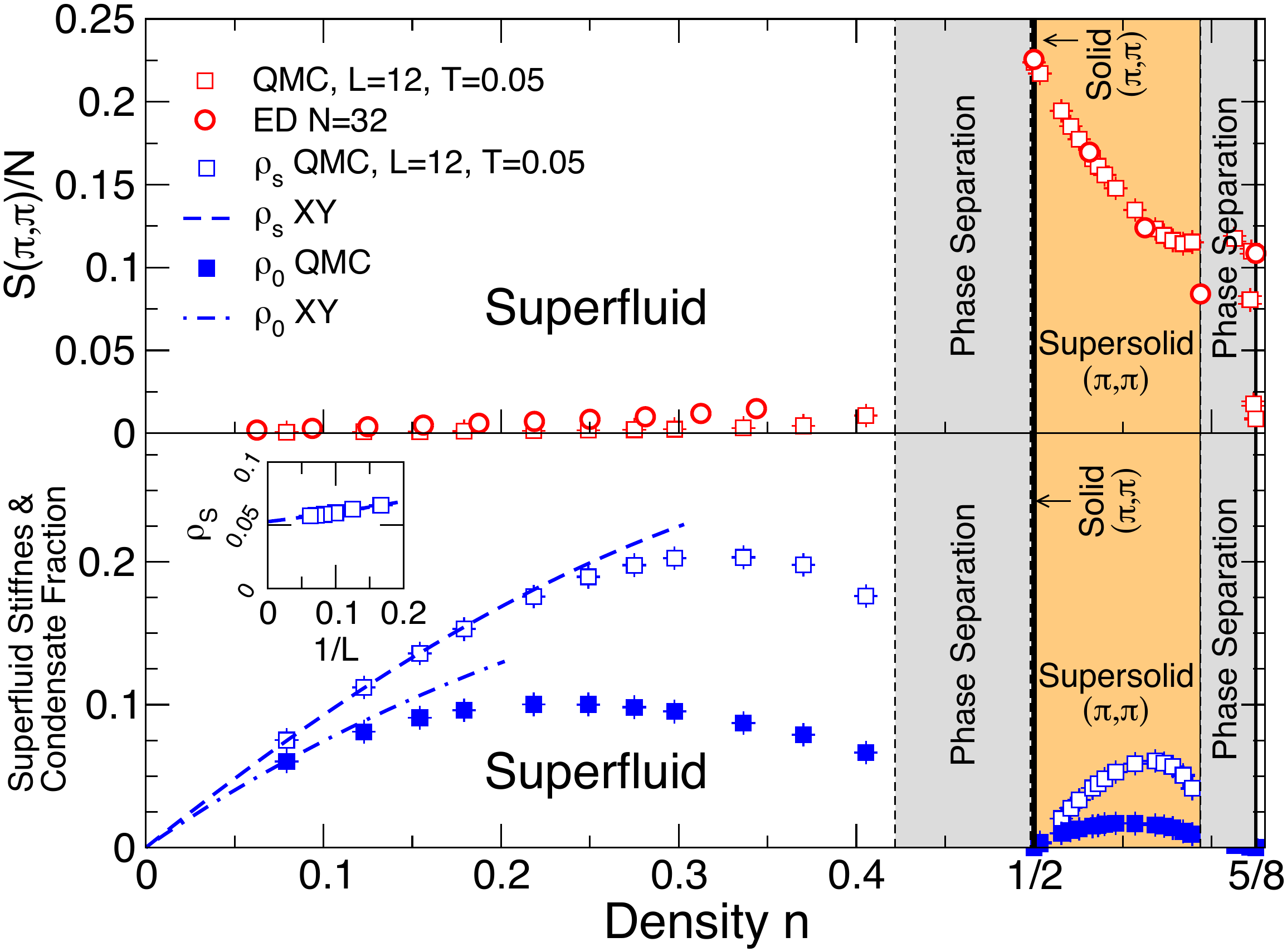}
    \caption{Upper graph: density $n$ as a function of the chemical potential $\mu$ for the minimal 
    model~(c.f. Fig.~\protect{\ref{fig:pd_truncated}}) at $t=0.5$ up to density $n=5/8$.
    Lower graph, upper panel: checkerboard order parameter $S(\pi,\pi)/N$ from ED  ($N=32$) and QMC ($N=12\times 12$) 
    simulations. Lower panel: superfluid stiffness $\rho_s$ and condensate fraction $\rho_0$ obtained by QMC. Inset: $\rho_s$
    at fixed $\beta=20$ and $\mu=15$ for $L=6,8,\dots, 16$.}
    \label{fig:qmc_ed}
\end{figure}

The numerical data for the density $n$ as function of $\mu$ is shown in the top panel of Fig.~\ref{fig:qmc_ed}. The systems starts to fill at 
$\mu_1=-4t=-2$ and the density behaves smoothly up to $\mu\approx 4.2$ where the system phase separates between a superfluid component at density $n\approx 0.4$ and the CS at $n=1/2$. Note that there is no $n=1/3$ plateau present at $t=0.5$. The $n=1/3$ plateau is recovered for $t=0.25$ (see below) implying that at least some differences to the SCA can be resolved in terms of a renormalized $t$.
At $\mu\approx 13.6$ the CS gets doped with 
particles which condense without destroying the solid and thus form a CSS. 
The supersolid remains stable up to $\mu\approx 16.3(1)$ where phase separation occurs anew, 
this time between the CSS and the $n=5/8$ solid. 

The phases just discussed are determined by measurements of the superfluid stiffness $\rho_s=\frac{1}{2\beta L^2} \langle W_x^2 + W_y^2 \rangle$ where $W_x$ and $W_y$ are the total winding numbers in $x$ and $y$ directions, the condensate fraction $\rho_0=\lim_{j\rightarrow\infty}\langle b^\dagger_j b^{\phantom{\dagger}}_0\rangle$
as well as the checkerboard charge order parameter $S(\pi,\pi)/N=\frac{1}{N^2}\sum_{i,j} (-1)^{i-j} \langle n_i n_j\rangle$ displayed in the central and bottom panels of Fig.~\ref{fig:qmc_ed}. 
At low densities $0\leq n \lesssim 0.4$ the
system is indeed superfluid, with a finite superfluid stiffness $\rho_s$ and condensate fraction $\rho_0$. At very low densities both quantities are
in very good quantitative agreement with the values obtained for noninteracting hardcore bosons~\cite{HardcoreXY}. At density $n=1/2$ the 
checkerboard order is highlighted by the structure factor data obtained with ED and QMC for different system sizes. 
As one dopes the CS with additional particles a sizable CSS emerges for $1/2<n\lesssim 0.59$, therefore confirming the SCA prediction. 
Finite size effects are small in the CSS phase, as witnessed by $S(\pi,\pi)/N$ being essentially unchanged from $N=32$ (ED) to 
$N=144$ (QMC), while the finite size extrapolation of the stiffness $\rho_s$ (QMC, inset) converges to a finite value, showing that the CSS is stable in the thermodynamic limit.

The stability of the CSS is surprising, since it is commonly believed that the CSS is unstable for
hardcore bosons with only nearest-neigbor hopping~\cite{BatrouniCheckerSupersolid}. In the present case it  can however be shown that
the instability towards domain-wall formation~\cite{sengu05} is absent, therefore providing an explanation for the stability of the CSS.
We remark that in the CSS phases the solid order is more pronounced than the superfluid component,
due to the vicinity of the solid. We therefore expect the supersolid to first give way to a solid phase by a Kosterlitz-Thouless transition, followed by an Ising transition to a normal bose liquid with increasing temperature~\cite{schmi08b}.

We have performed ED simulations for $t=0.25$, 0.75 and 1 to further check for the presence of phases predicted by the SCA~\cite{unpublished}.
At $t=0.25$ we find evidence for a $n=1/3$ plateaux, a CSS {\em below} as well as above the $n=1/2$ CS.
At $t=0.75$ the CSS above the $n=1/2$ plateau and the phase separation below the solid are reduced in density extent, compared to $t=0.5$.
Finally at $t=1$, using QMC and ED, we find a $n=1/2$ and $n=2/3$ plateau, while the CSS above $n=1/2$ is tiny if present at all. The remaining regions are superfluid. The phase transitions between the superfluid and the solids at $t=1$ deserve further study.\\
Although the SCA approach does not treat the quantum fluctuations quantitatively correctly, our numerical investigations confirms
 that many of the qualitative SCA phases are correct and indeed present. 
Based on this validation we now address the effect of the finite range approximation.

\begin{figure}
    \begin{center}
        \includegraphics*[width=0.87\columnwidth]{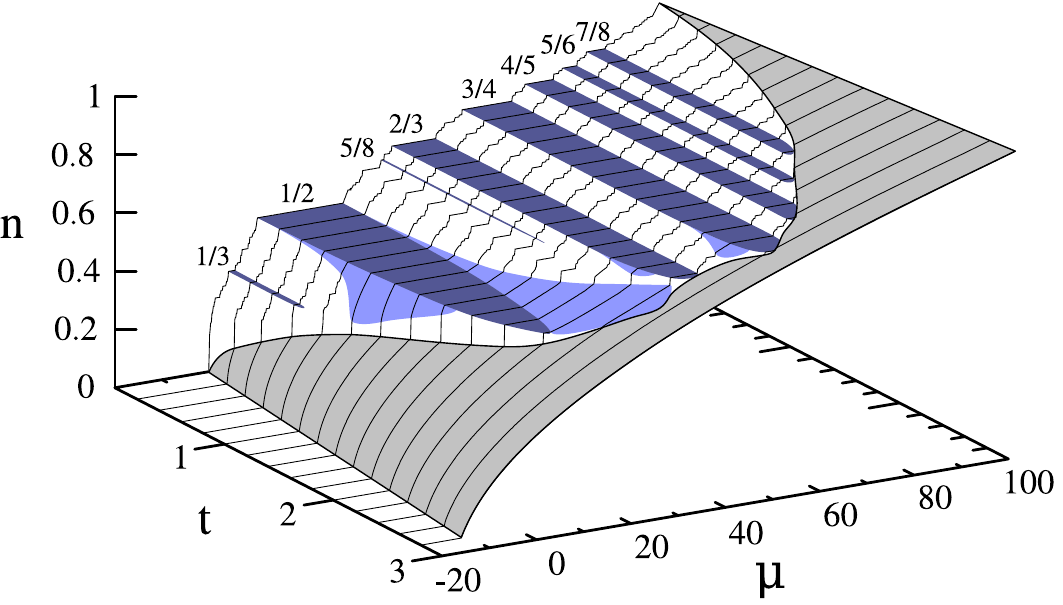}
    \end{center}
    \caption{(Color online) 
    Phase diagram obtained by SCA for different values of $t$ including all 3B terms with amplitudes $W_{ijk}>0.001$.
    Grey regions are superfluid, and light (dark) blue denotes supersolids (solids), which were already present in the
    minimal model c.f. Fig.~\protect{\ref{fig:pd_truncated}. The white regions encompass phase separation and further solids and
    supersolids.}
  \label{fig:sca_barier_0.001}}
\end{figure}

\paragraph{Effect of longer range interactions}

The numerical solution of the Hamiltonian~(\ref{eq:hamilton}) including the 3B interactions at all distances is a formidable task.
Even the classical problem for $t=0$ is non-trivial due to the discrete structure imposed by the square lattice.
Here we are merely interested whether the generic features found in the truncated model are stable upon the inclusion of the long-range nature of the 3B interactions.  To this end we use the SCA including all 3B interactions in the Hamiltonian with amplitudes $W_{ijk}>0.001$.
Minimizations are done on all clusters up to 24 sites. While these clusters might still be too small to represent all structures
at small $t$, they are amply sufficient to confirm that many of the solids and supersolids present in the minimal model survive the inclusion of the
long-range 3B couplings. The phase diagram is shown in Fig.~\ref{fig:sca_barier_0.001}. Most importantly we confirm the stability of
all plateaux of the minimal model, plus the supersolids below and above the CS, as well as supersolids below the
$n=2/3$ and $3/4$ solids. While in some solids at large densities (e.g. $n=5/6$ and $7/8$) the degeneracy on the classical level is lifted, we find that other solids ($n=5/8$ and $3/4$) change slightly the charge order pattern. This reflects the general tendency of the 3B interactions at large densities to form a triangular Wigner crystal of holes. The optical lattice in the square geometry then acts as an incommensurate substrate, possibly giving rise to physics similar to the Frenkel-Kontorova model and to slow equilibration (glassiness) 
due to many almost degenerate charge configurations.

\paragraph{Conclusion}
We studied a model of hardcore bosons on the square lattice interacting solely by slowly decaying 3B interactions, which is directly relevant for future experiments on ultracold gases of polar molecules confined to optical lattices. We find a rich phase diagram consisting of many solid, superfluid, and supersolid phases. The long-range nature of the 3B interactions results in a zoo of crystalline phases in the limit of small $t$. The large number of competing states will probably also lead to difficulties in the equilibriation of the experimental system in this limit. 

The most important finding of our work is that a system which only contains 3B interactions realizes supersolid phases. Extended supersolid phases exist around the $n=1/2$ CS. Our findings therefore suggest that ultracold gases of polar molecules confined to optical lattices are promising candidates to observe for the first time a supersolid on a lattice in experiment. 

\begin{acknowledgments}
KPS acknowledge ESF and EuroHorcs for financial support through his EURYI.  
JD and AML acknowledge support from the SNF and MaNEP. Part of the numerical simulations have been
performed at CSCS (Manno).
\end{acknowledgments}

\end{document}